\documentclass[aps,twocolumn,groupedaddress,showpacs]{revtex4}

\usepackage{graphicx}
\usepackage{epsfig}

\begin{document}

\bibliographystyle{apsrev}

%\preprint{}

\title{Appearance of fractional charge in the noise of non-chiral
Luttinger liquids}

\author{Bj{\"o}rn~Trauzettel$^1$, In{\`e}s~Safi$^1$, Fabrizio~Dolcini$^2$,
and Hermann~Grabert$^2$}

\affiliation{${}^1$Laboratoire de Physique des Solides, Universit\'e Paris-Sud,
91405 Orsay, France\\
${}^2$Physikalisches Institut, Albert-Ludwigs-Universit\"at, 79104
Freiburg, Germany}

\date{\today}

\begin{abstract}
The current noise of a voltage biased interacting quantum wire
adiabatically connected to metallic leads is computed in presence
of an impurity in the wire. We find that in the weak
backscattering limit the Fano factor characterizing the ratio
between noise and backscattered current crucially depends on the
noise frequency $\omega$ relative to the ballistic frequency
$v_F/gL$, where $v_F$ is the Fermi velocity, $g$ the Luttinger
liquid interaction parameter, and $L$ the length of the wire. In
contrast to chiral Luttinger liquids the noise is not only due to
the Poissonian backscattering of fractionally charged
quasiparticles at the impurity, but also depends on Andreev-type
reflections at the contacts, so that the frequency dependence of
the noise needs to be analyzed to extract the fractional charge
 $e^*=e g$ of the bulk excitations.

\end{abstract}
\pacs{71.10.Pm, 72.10.-d, 72.70.+m, 73.23.-b}

\maketitle

Shot noise measurements are a powerful tool to observe the charge
of  elementary excitations of interacting electron systems. This
is due to the fact that in the Poissonian limit of uncorrelated
backscattering of quasiparticles from a weak impurity, the low
frequency current noise is directly proportional to the
backscattered charge \cite{blante00}. This property turns out to
be particularly
useful in probing the fractional charge of excitations in
one-dimensional (1D) electronic systems, where correlation effects
destroy the Landau quasiparticle picture and give rise to collective
excitations, which in general obey unconventional statistics, and
which have a charge different from the charge $e$ of an electron \cite{pham00}.
In particular, for fractional quantum Hall (FQH) edge state devices,
which at filling fraction $\nu=1/m$ ($m$ odd integer) are usually
described by the {\it chiral} Luttinger liquid (LL) model, it has
been predicted that shot noise should allow for an observation of
the fractional charge $e^*=e\nu$ of backscattered Laughlin
quasiparticles \cite{kane94}. Indeed, measurements at $\nu=1/3$ by
two groups \cite{samin97,depic97} have essentially confirmed this
picture. The question arises whether similar results can be expected also
for {\it non-chiral} LLs, which are believed to be realized in carbon
nanotubes \cite{bockr99} and single channel semiconductor quantum
wires \cite{yacob96}. Although a non-chiral LL can be modelled
through the very same formalism as a pair of chiral LLs, some
important differences between these two kinds of LL systems have
to be emphasized. In particular, in chiral LL devices right- and
left-moving charge excitations are spatially separated, so that
their chemical potentials can be independently tuned in a
multi-terminal Hall bar geometry. In contrast, in non-chiral LL
systems, right- and left-movers are confined to the same channel,
and it is only possible to control the chemical potentials of the
Fermi liquid reservoirs attached to the 1D wire. This in turn
affects the chemical potentials of the right- and left-moving
charge excitations in a non-trivial way depending on the
interaction strength, and implies crucial differences between
chiral and non-chiral LLs, for instance, the conductance
in the former case depends on the LL parameter $g=\nu$ \cite{kane92},
while in the latter case it is independent of $g$
\cite{safi95,inhom95,safi97}. Hence, the
predictions on shot noise properties of FQH systems are not
straightforwardly generalizable to the case of non-chiral LLs,
which therefore deserve a specific investigation. Previous
theoretical calculations of the shot noise of non-chiral LL
systems have shown that, even in the weak backscattering limit,
the zero frequency noise of a finite-size non-chiral LL does not
contain any information about the fractional charge backscattered
off an impurity \cite{ponom99,trauz02}, but is rather proportional
to the charge of an electron. This result, as well as the above
mentioned interaction independent DC conductance, prevents easy
access to the interaction parameter $g$.

On the other hand, a quantum wire behaves
as a Andreev-type resonator for an incident electron,
which gets transmitted as series of current spikes 
\cite{safi95}. The reflections of charge excitations 
at both contacts are called Andreev-type reflections 
because they are momentum
conserving as ordinary Andreev reflections \cite{safi95,sandl98}. 
Since the transmission dynamics in the Andreev-type resonator 
depends on $g$, finite frequency
 transport can resolve internal properties of the wire.
This is, in fact, the case for the AC conductance
\cite{safi95,safi97,blant98}. However, finite
frequency conductance measurements are limited in
the AC frequency range since the frequency must be low enough to ensure
quasi-equilibrium states in the reservoirs in order to compare
experiments to existing theories. The better alternative
is to apply a DC voltage and measure finite frequency
current noise.
Here, exploring the out of equilibrium regime, it is shown that
the noise as a function of frequency has a periodic structure
with period $2\pi\omega_L$, where $\omega_L=v_F/gL$ is the inverse of
the traversal time of a charge excitation with plasmon velocity $v_F/g$
through the wire of length $L$. The Fano factor
oscillates and we will show that by averaging over
$2\pi\omega_L$, the effective charge $e^*=e g$ can be extracted
from noise data.

In order to analyze the noise of non-chiral LLs it is essential to
study the inhomogeneous LL (ILL) model \cite{safi95,inhom95},
which takes the finite length of the interacting wire and the
coupling to the reservoirs explicitly into
account. This model is governed by the Hamiltonian
${\mathcal{H}} ={\mathcal{H}}_{0}  + {\mathcal{H}}_{B} +
{\mathcal{H}}_{V}$ ,
where ${\mathcal{H}}_{0}$ describes the interacting wire, the
leads and their mutual contacts, ${\mathcal{H}}_{B}$ accounts for
the electron-impurity interaction, and ${\mathcal{H}}_{V}$
represents the coupling to the electrochemical bias applied to
the wire. Explicitly, the three parts of the Hamiltonian read
\begin{eqnarray}
{\mathcal{H}}_0 &=&\frac{\hbar v_F}{2}  \int_{-\infty}^{\infty}
 dx \left[ \Pi^2 + \frac{1}{g^2(x)}
(\partial _x\Phi )^2\right]  \, , \label{L0}  \\
{\mathcal{H}}_B &=& \lambda \cos{[\sqrt{4 \pi} \Phi(x_0,t)+2 k_F
x_0]} \label{LB} \; ,\\
{\mathcal{H}}_{V}  &=&    \int_{-\infty}^{\infty}
\frac{dx}{\sqrt{\pi}} \, \mu(x) \,
\partial_x \Phi(x,t) \; . \label{LV}
\end{eqnarray}
Here, $\Phi(x,t)$ is the standard Bose field operator in
bosonization and $\Pi(x,t)$ its conjugate momentum density \cite{gogol98}.
The Hamiltonian ${\mathcal{H}}_0$
describes the (spinless) ILL, which is known to capture the essential
physics of a quantum wire  adiabatically connected to metallic
leads. The interaction parameter $g(x)$ is space-dependent and its
value is 1 in the bulk of the non-interacting leads
 and $g$ in the bulk of the wire ($0 < g < 1$ corresponding to repulsive
interactions). The variation of $g(x)$ at the contacts from 1 to
$g$ is assumed to be smooth, i.e. to occur within a characteristic
length $L_s$ fulfilling $\lambda_F \ll L_s \ll L$, where
$\lambda_F$ is the electron Fermi wavelength. Since the
specific form of the function $g(x)$ in the contact region will
not influence physical features up to energy scales of order
$\hbar v_F/L_s$,  we shall, as usual, adopt a step-like function.
The Hamiltonian ${\mathcal{H}}_B$ is the dominant $2k_F$ backscattering term
at the impurity site $x_0$, and introduces a strong non-linearity
in the field $\Phi$. Finally, Eq.~(\ref{LV}) contains the applied
voltage. In most experiments
leads are normal 2D or 3D contacts, i.~e. Fermi liquids. However, since
we are interested in properties of the {\it wire}, a detailed
description of the leads would in fact be superfluous. One can
 account for their main effect, the applied bias voltage at the
contacts, by treating them as non-interacting 1D systems ($g=1$). 
The only essential properties originating from
the Coulomb interaction that one needs to retain are (i) the
possibility to shift the band-bottom of the leads, and (ii)
electroneutrality \cite{trauz02}. Therefore, the function $\mu(x)$
appearing in Eq.~(\ref{LV}), which describes the externally
tunable electrochemical bias, is taken as piecewise constant
$\mu(x<-L/2) = \mu_L$, $\mu(x>L/2) = \mu_R$ corresponding to an
applied voltage $V = (\mu_L-\mu_R)/e$. In contrast, the QW itself
does not remain electroneutral in presence of an applied voltage,
and its electrostatics emerges naturally from
Eqs.~(\ref{L0})-(\ref{LV}) with $\mu=0$ for $|x|<L/2$
\cite{safi97,egger96}.

In bosonization, the current operator is related to the Bose
field $\Phi$ through $j (x,t) = - (e/\sqrt{\pi}) \partial_t \Phi(x,t)$.
Moreover, the finite frequency noise is defined as
\begin{eqnarray} \label{noise}
S(x,y;\omega) = \int_{-\infty}^{\infty} dt e^{i\omega t}
\left\langle \left\{ \Delta j (x,t) , \Delta j(y,0) \right\}
\right\rangle \; ,
\end{eqnarray}
where $\{ \, , \, \}$ denotes the anticommutator and $\Delta j(x,t) =
j(x,t) - \langle j(x,t) \rangle$ is the current fluctuation
operator. Since we investigate non-equilibrium properties of
the system, the actual calculation of the averages of current and
noise are performed within the Keldysh formalism \cite{keldy64}.

The average current $I \equiv \langle j(x,t) \rangle$ can be
expressed as $I=I_0 -I_{\rm BS}$, where $I_0=(e^2/h) V$ is the current in
the absence of an impurity, and $I_{\rm BS}$ is the backscattered
current. For arbitrary impurity strength, temperature, and
voltage, the backscattered current can be written in the compact
form
\begin{equation} \label{BScurrent}
I_{\rm BS}(x,t) = -\frac{\hbar \sqrt{\pi}}{e^2}
\int_{-\infty}^\infty dt' \sigma_0(x,t;x_0,t') \langle j_B
(x_0,t') \rangle_\rightarrow \ ,
\end{equation}
where $\sigma_0(x,t;x_0,t')$ is the non-local conductivity of the clean wire
derived in \cite{safi95,safi97,blant98}. 
In Eq.~(\ref{BScurrent}), we have introduced the ``backscattered current operator''
\begin{equation} \label{BSoperator}
j_B(x_0,t) \equiv - \frac{e}{\hbar} \frac{\delta
{\mathcal{H}}_B}{\delta \Phi(x_0,t)} (\Phi + A_0) \ ,
\end{equation}
where $A_0(x_0,t)$ is a shift of the phase field emerging when one
gauges away the applied voltage. For a DC voltage this shift
simply reads $A_0(x_0,t)=\omega_0 t/2\sqrt{\pi}$ with $\omega_0 =
eV/\hbar$ and $I_{\rm BS}$ does not depend on $x$ and $t$.
Furthermore, we have introduced a ``shifted average''
$\langle \dots \rangle_\rightarrow$, which is evaluated with
respect to the shifted Hamiltonian
${\mathcal H}_\rightarrow = {\mathcal{H}}_0[\Phi] + {\mathcal{H}}_B[\Phi+A_0]$.
A straightforward though
lengthy calculation shows that the finite frequency current noise
(\ref{noise}) can (again for arbitrary impurity strength,
temperature, and voltage) be written as the sum of three contributions
\begin{equation} \label{ff_result}
S(x,y;\omega)=S_0(x,y;\omega)+S_A(x,y;\omega)+S_C(x,y;\omega) \; .
\end{equation}
The first part of Eq.~(\ref{ff_result}), $S_0(x,y;\omega)$, is the
current noise in the absence of a backscatterer, and can be
related to the conductivity $\sigma_0(x,y;\omega)$ by the
fluctuation dissipation theorem \cite{ponom96}
\begin{equation}
S_0(x,y;\omega)= 2 \hbar \omega \coth\left( \frac{\hbar \omega}{2 k_B T}
\right) \Re [\sigma_0(x,y;\omega)] \; .
\label{s0thermal}
\end{equation}
The conductivity can be expressed by the Kubo formula
$\sigma_0(x,y;\omega) =  2 (e^2/h) \omega C_0^R(x,y;\omega)$, where
\[
C_0^R(x,y;\omega) =  \int_0^\infty dt e^{i \omega t} \langle [
\Phi(x,t), \Phi(y,0)] \rangle_0
\]
is the time-retarded correlator of the equilibrium ILL model
in the absence of an impurity. It is important to note
that usually the relation (\ref{s0thermal}) is only valid in thermal
equilibrium, and the Kubo formula is based on linear response theory.
However, due to the fact that in the absence of an impurity the
current of a quantum wire attached to Fermi liquid reservoirs is
linear in the applied voltage \cite{safi95,safi97},
Eq.~(\ref{s0thermal}) is also valid out of equilibrium.

The other two terms in Eq.~(\ref{ff_result}) arise from the
partitioning of the current at the impurity site. The second term
is related to the anticommutator of the backscattered current
operator $j_B$, and reads
\begin{eqnarray} \label{sb}
&& S_A(x,y;\omega) = \\
&& \frac{1}{\pi} \left(\frac{h}{2 e^2}\right)^2
\sigma_0(x,x_0;\omega) f_A(x_0,\omega) \sigma_0(x_0,y;-\omega)
\nonumber
\end{eqnarray}
with
\[
f_A(x_0,\omega) = \int_{-\infty}^\infty dt \, e^{i \omega t}
\left\langle \left\{ \Delta j_B(x_0,t), \Delta j_B(x_0,0)
\right\} \right\rangle_\rightarrow \; ,
\]
where $\Delta j_B(x,t) \equiv j_B(x,t) - \langle j_B(x,t) \rangle_\rightarrow$.
Finally, the third part of Eq.~(\ref{ff_result})
is related to the time-retarded commutator of $j_B$ and can
be expressed as
\begin{eqnarray} \label{sm2}
&& S_C(x,y;\omega) = \\
&& \frac{h}{2 e^4 \omega} \Bigl\{ S_0(x,x_0;\omega)
f_C(x_0,-\omega) \sigma_0(x_0,y;-\omega) \nonumber \\
&& - S_0(y,x_0;-\omega) f_C(x_0,\omega) \sigma_0(x_0,x;\omega)
\Bigr\} \nonumber
\end{eqnarray}
with
\[
f_C(x_0,\omega) = \int_0^\infty dt \left( e^{i \omega t}-1 \right)
\left\langle \left[ j_B(x_0,t),j_B(x_0,0) \right]
\right\rangle_\rightarrow \; .
\]
The fractional charge is expected to emerge only in the limit of
weak backscattering through the ratio between shot noise and
backscattered current. We thus focus on the case of a weak
impurity, retaining in the expressions (\ref{BScurrent}) and
(\ref{ff_result}) only contributions of second order in the
impurity strength $\lambda$. Furthermore, we concentrate on the
shot noise limit of large applied voltage.

The backscattering current (\ref{BScurrent}) may be written as
$I_{\rm BS} = (e^2/h) {\cal R} V $, where ${\cal R}$ is an
effective reflection coefficient. Contrary to a non-interacting
electron system, ${\cal R}$ depends   on voltage and interaction
strength \cite{kane92,dolci03}. In the weak backscattering limit
${\cal R} \ll 1$, and its actual value can readily be determined
from a   measurement of the current voltage characteristics.
Importantly, for temperatures in the window $eV {\cal R} \gg k_B T
\gg \{ \hbar\omega, \hbar \omega_L \}$ the noise can be shown to
be dominated by the second term in Eq.(\ref{ff_result}) and to
take the simple form
\begin{equation} \label{excessnoise}
S (x,x;\omega) \simeq  2 e F(\omega) I_{\rm BS} \; ,
\end{equation}
where $x=y$ is the point of measurement (in either of the two
leads). In Eq.~(\ref{excessnoise}), the contributions neglected
 are of order $k_BT/eV{\cal R}$. The Fano
factor
\begin{equation}
F(\omega)= \frac{h^2}{e^4} |\sigma_0(x,x_0;\omega)|^2
\label{Fano-cond}
\end{equation}
is given in terms of the non-local conductivity
$\sigma_0(x,x_0;\omega)$ relating the measurement point $x$ to the
impurity position $x_0$, and reads explicitly
\begin{eqnarray} \label{Fano-fun}
F(\omega) =   ( 1-\gamma)^2 \frac{1+\gamma^2+2\gamma \cos \left(
\frac{2\omega (\xi_0+1/2)}{\omega_L} \right)}{1+\gamma^4-2\gamma^2
\cos \left( \frac{2 \omega}{\omega_L}\right)} \; .
\end{eqnarray}
The latter expression is, in fact, independent of the point of
measurement $x$ and of temperature. On the other hand, it depends,
apart from the frequency $\omega$, on the (relative) impurity
position $\xi_0=x_0/L$, and the interaction strength through
$\gamma = (1-g)/(1+g)$.
%%%%%%%%%%%%%%%%%%%%%%%%%%%%%%%%%%%%%%%
%%%%%%%%%%%%%%%%%%%%%%%%%%%%%%%%%%%%%%%
%%%%%      FIGURE    1          %%%%%%
%%%%%%%%%%%%%%%%%%%%%%%%%%%%%%%%%%%%%%%
%%%%%%%%%%%%%%%%%%%%%%%%%%%%%%%%%%%%%%%
\begin{figure}
\vspace{0.3cm}
\begin{center}
\epsfig{file=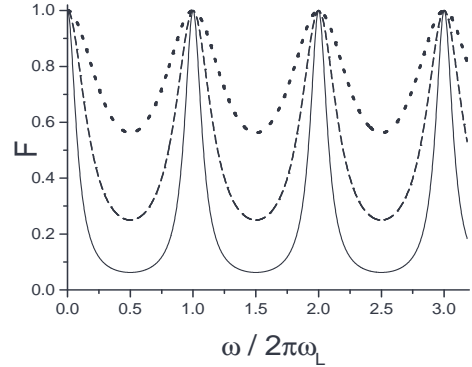,width=7cm,height=5.8cm}
\caption{\label{Fano_x000} The periodic function $F(\omega)$,
which determines the Fano factor, is shown as a function of
$\omega/2 \pi\omega_L$, for the case of an impurity at the
center of the wire ($x_0=0$) and three different values of the
interaction strength: $g=0.25$ (solid), $g=0.50$ (dashed), and
$g=0.75$ (dotted). In the regime $\omega/\omega_L \ll 1$, the
function tends to 1 independent of the value of $g$, but for
$\omega \lesssim \omega_L$ the curve strongly depends on the
interaction parameter $g$. In particular, $g$ can be obtained as
the average over one period.}
\end{center}
\end{figure}

The central result (\ref{excessnoise}) shows that the ratio
between the shot noise and the backscattered current crucially
depends on the frequency regime one explores. In particular, for
$\omega \rightarrow 0$, the function $F$ tends to 1, independent
of the value of the interaction strength. Therefore, in the regime
$\omega \ll \omega_L$ the observed charge is just the electron
charge. In contrast, at frequencies comparable to $\omega_L$ the
behavior of $F$ as a function of $\omega$ strongly depends on the
LL interaction parameter $g$, and signatures of LL physics emerge.
This is shown in Fig.~\ref{Fano_x000} for the case of an impurity
located at the center of the wire. Then, $F(\omega)$ is periodic,
and the value at the minima coincides with $g^2$. Importantly, $g$
is also the mean value of $F$ averaged over one period $2\pi
\omega_L$,
\begin{equation} \label{s_ex_average}
\left\langle S (x,x;\omega) \right\rangle_{\omega} \equiv
\frac{1}{2 \pi \omega_L} \int_{-\pi \omega_L}^{\pi \omega_L}
S (x,x;\omega) \simeq 2 e g I_{\rm BS} \; ,
\end{equation}
where again terms of order $k_BT/eV{\cal R}$ are neglected.
Seemingly, Eq.~(\ref{s_ex_average}) suggests that quasiparticles
with a fractional charge $e^*=eg$ are backscattered off the
impurity in the quantum wire.

Let us discuss the physical origin of this appearance of the
fractional charge. We first consider the case of an infinitely
long quantum wire. In the limit $L\to\infty$, i.e.\ $\omega_L\to
0$, $\xi_0 \rightarrow 0$, the function $F(\omega)$ becomes
rapidly oscillating and its average over any finite frequency
interval approaches $g$. Hence, we recover in this limit the
result for the homogeneous LL system \cite{kane94}, where the shot
noise is directly proportional to the fractional charge $e^*=ge$
backscattered off the impurity. However, as shown above, the value
of the fractional charge $e^*$ can be extracted not only in the
borderline case $\omega\gg\omega_L$, but already for frequencies
$\omega$ of order $\omega_L$. This is due to the fact that,
although the contacts are adiabatic, the mismatch between
electronic excitations in the leads and in the wire inhibits the
direct penetration of electrons from the leads into the wire;
rather a current pulse is decomposed into a sequence of fragments
by means of Andreev-type reflections at the contacts
\cite{safi95}. These reflections are governed by the coefficient
$\gamma=(1-g)/(1+g)$, which depends on the interaction strength.
The zero frequency noise is only sensitive to the sum of all
current fragments, which add up to the initial current pulse
carrying the charge $e$. However, when $2\pi/\omega$ becomes
comparable to the time needed by a plasmon to travel
from the contact to the impurity site, the noise resolves the
current fragmentation at the contacts. The sequence of
Andreev-type processes is encoded in the non-local conductivity
$\sigma_0(x,x_0;\omega)$ relating the measurement point $x$ and
the impurity position $x_0$. This enters into the Fano factor
(\ref{Fano-cond}) and allows for an identification of $e^*$ from
finite frequency noise data.
%%%%%%%%%%%%%%%%%%%%%%%%%%%%%%%%%%%%%%%
%%%%%%%%%%%%%%%%%%%%%%%%%%%%%%%%%%%%%%%
%%%%%      FIGURE    2          %%%%%%
%%%%%%%%%%%%%%%%%%%%%%%%%%%%%%%%%%%%%%%
%%%%%%%%%%%%%%%%%%%%%%%%%%%%%%%%%%%%%%%
\begin{figure}
\vspace{0.3cm}
\begin{center}
\epsfig{file=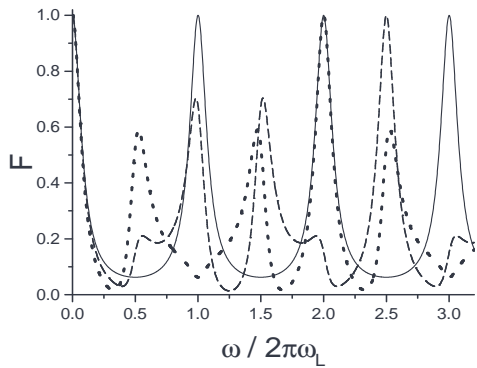,width=7cm,height=5.8cm,clip=}
\caption{\label{Fano_g025} The Fano factor $F(\omega)$ is shown for
the interaction strength $g=0.25$ and three different values of the
(relative) impurity position $\xi_0=x_0/L$: $\xi_0=0$ (solid),
$\xi_0=0.10$ (dashed), and $\xi_0=0.25$ (dotted).}
\end{center}
\end{figure}

When the impurity is located away from the center of the wire,
$F(\omega)$ is no longer strictly periodic, as shown in Fig.
\ref{Fano_g025}. In that case, the combined effect of Coulomb
interactions and an off-centered impurity can lead to a very
pronounced reduction of the Fano factor for certain noise
frequencies (see Fig.~\ref{Fano_g025}). Moreover, even if the impurity is 
off-centered, the detailed predictions
(\ref{excessnoise}) and (\ref{Fano-fun}) should allow to
gain valuable information on the interaction constant $g$ from
the low frequency behavior of the Fano factor determined by
\[
F(\omega) = 1-(1-g^2) (1+4 g^2 \xi_0(1+\xi_0))
\left(\frac{\omega L}{2 v_F}\right)^2 + 
\dots \; .
\]
The latter expression is valid in the
parameter regime $eV{\cal R}  \gg k_B T \gg \hbar \omega_L \gg
\hbar \omega$. 

In conclusion, the appearance of fractional charge $e^*=eg$
 in the finite frequency noise of non-chiral LLs is due to a
combined effect of backscattering of  bulk quasiparticles at the
impurity  and of  Andreev-type reflections of plasmons at the
interfaces of wire and leads. The fractional charge $e^*$
 can be extracted from the noise by
averaging it over a frequency range $[-\pi \omega_L,\pi\omega_L]$
in the out of equilibrium regime. For single-wall carbon nanotubes
we know that $g \approx 0.25$, $v_F \approx 10^5 \; {\rm m/s}$,
and their length can be up to 10 microns. Thus, we estimate
$\pi\omega_L \approx 100 \; {\rm GHz} \dots 1 \; {\rm THz}$, which
is a frequency range that seems to be experimentally accessible
\cite{schoe97,deblo03}. Moreover, the requirement $eV \gg \hbar
\omega_L$ should be fulfilled in such systems for $eV \approx 10
\dots 50 \; {\rm meV}$, a value which is well below the subband
energy separation of about $1 \; {\rm eV}$.

We thank H.~Bouchiat, R.~Deblock, R.~Egger, D.~C.~Glattli, and P.
Roche for interesting discussions. Financial support by the EU
is gratefully acknowledged.

\end{document}